\documentclass[12pt]{iopart}
\usepackage{iopams}
\eqnobysec
\newcommand{\be}{\begin{equation}}
\newcommand{\ee}{\end{equation}}

\newcommand{\ba}{\begin{array}}
\newcommand{\ea}{\end{array}}
\newcommand{\bea}{\begin{eqnarray}}
\newcommand{\eea}{\end{eqnarray}}
\newcommand{\qed}{\begin{flushright} $\square$
                  \end{flushright}
}
\newcommand{\acb}{\alpha, \beta}
\newcommand{\apb}{\alpha+ \beta}
\newcommand{\nn}{\nonumber \\}
\newcommand{\bel}{\begin{equation}\label}
\newcommand{\pab}{P_n^{(\acb)}(x)}

\newtheorem{theorem}{Theorem}

\newtheorem{lemma}{Lemma}

\begin{document}

\title{An infinite family of superintegrable Hamiltonians with reflection in the plane}
\author{Sarah Post$^1$, Luc Vinet$^1$ and Alexei Zhedanov $^2$  }
\address{$^1$ Centre de Recherches Math\'ematiques. Universit\'e de Montr\'eal. Montr\'eal CP6128 (QC) H3C 3J7, Canada}
\address{$^2$ Donetsk Institute for Physics and Technology. Donetsk 83114, Ukraine}
\ead{post@crm.umontreal.ca, luc.vinet@umontreal.ca}
\begin{abstract} We introduce a new infinite class of superintegrable quantum systems in the plane. Their Hamiltonians involve reflection operators. The associated Schr\"odinger equations admit separation of variables in polar coordinates and are exactly solvable. The angular part of the wave function is expressed in terms of little -1 Jacobi polynomials. The spectra exhibit "accidental" degeneracies. The superintegrability of the model is proved using the recurrence relation approach. The (higher-order) constants of motion are constructed and the structure equations of the symmetry algebra obtained. 
\end{abstract}
%\maketitle

\section{Introduction}
A quantum system in n-dimensions is integrable if it admits $n$ integrals of motion that are in involution. It is superintegrable if it possesses additional constants of the motion, i.e. more than there are degrees of freedom. In such cases, only subsets of $n$ of the symmetry generators can commute amongst themselves. A model is called maximally superintegrable if it has $2n-1$ conserved quantities, including the Hamiltonian, that are algebraically independent. 

In view of the special properties and applications that superintegrable models enjoy, there is considerable interest in enlarging the set of documented systems in that category. While not doing justice to the many contributors to this field, let us mention in this connection that the recent identification of the TTW family \cite{1} in particular, has attracted a lot of attention. It stimulated much effort aimed at demonstrating the superintegrability of (the higher-order) members of the family \cite{2}, obtaining the constants of the motion and determining the structure of the symmetry algebras \cite{3, Marquette}. To that end, the recurrence approach \cite{4}  proved most general and powerful  for determining the explicit integrals of the quantum systems. 

We here wish to add to this body of knowledge by introducing another infinite family of maximally superintegrable systems in the plane. This new family shares many features with the TTW one; it has however a distinctive trait in that its Hamiltonians involve a reflection operator. 

Hamiltonians with reflection symmetry are customary in the context of the Calogero-Sutherland integrable models and their generalizations. The exchange operator formalism is indeed best suited to express the conserved quantities of these systems, to show that they are in involution and to find the ladder operators \cite{5,6,7}. Interestingly, it is precisely this formalism that has been used in \cite{2} to address the superintegrability of the TTW system. It should also be recalled that the creation and annihilation operators of the parabose oscillator are realized with reflection operators \cite{8,9,10}. This background provides reasons to study problems with reflection operators in addition to the intrinsic interest that finding new superintegrable models entails. 

The increasing appreciation that certain systems require the use of reflection operators for their description \cite{11,12, 13} is not unrelated to the fact that classical orthogonal polynomials that are eigenfunctions of first-order differential-difference operators with reflections have recently been discovered \cite{14,15,16, 17} . The simplest among those are the little -1 Jacobi polynomials which turn out to be $q \rightarrow -1$ limits of the little q-Jacobi polynomials \cite{14} . The identification and characterization of these new classes of polynomials is likely to enlarge the set of problems that are deemed exactly-solvable. 

We have in fact already found this to be the case in a recent article \cite{18} where we have considered a realization of supersymmetric quantum mechanics, that is achieved by using as supercharges such differential-difference operators with reflections. We have obtained through this approach an exactly solvable supersymmetric extension of the Scarf I Hamiltonian and found that its wave functions are given in terms of little -1 Jacobi polynomials. As it turns out, this system will be a key building block in the construction of our superintegrable family since it will essentially be the angular part of our 2-d Hamiltonians and play the role of the constant of motion responsible for separation of variables in polar coordinates. As we soon shall see, the Hamiltonians of the superintegrable models in our set, are formed by combining the radial part of the 2-d harmonic oscillator with this new, exactly-solvable, supersymmetric Hamiltonians in 1-d. We surmise that this illustrates a systematic way to construct problems that are exactly solvable and potentially superintegrable. 

The outline of the paper is as follows. In section 2, we introduce the family of Hamiltonians that is the object of this article. It comprises an infinite number of systems labeled by a real constant $k$ and further involves 3 real parameters $\omega, \alpha$ and $\beta.$ The exact solutions of the associated Schr\"odinger equations are provided in section 3. As already indicated, the little -1 Jacobi polynomials will appear.  For rational $k$, the energy spectra will show higher degeneracies, signaling the superintegrability of these systems which will be proven in section 4 using the recurrence approach and proven directly in the appendix. The (higher) integrals of motion will be explicitly obtained and the structure relations of the symmetry algebra will be determined. In section 5, we shall indicate how another infinite family of superintegrable models can be obtained from the one under consideration by trading, via coupling constant metamorphosis,  the harmonic potential  for the Coulomb one in the radial part \cite{19}.

\section{The infinite family of Hamiltonians}
We shall consider the following ensemble of Hamiltonians in $\mathbb{R}^2$ written in polar coordinates $(r, \theta):$
\bel{21}\fl H_k(r, \theta; \omega, \alpha, \beta)=-\partial _r^2-\frac1r\partial _r-\frac 1{r^2}\partial_\theta^2 +\omega^2 r+\frac{\alpha k^2}{2}\left(\frac{\frac\alpha 2-\cos k\theta\ R}{r^2\sin^2k\theta}\right)+\frac{\beta k^2}{2}\left(\frac{\frac\beta 2-\sin k\theta }{r^2\cos^2k\theta}\right)\ee
where $R$ is the reflection operators with respect to $\theta$
\bel{22} Rf(\theta)=f(-\theta).\ee
We require the following restrictions on the real parameters $\alpha, \beta, $ and $k$: $\alpha >-1, \beta > -1, $ $k \ne 0. $ We must also have $r \in [0, \infty)$ and $-\frac \pi{2k} \leq \theta \leq \frac{\pi}{2k}.$ 

The Hamiltonians $H_k(r, \theta; \omega, 0, \beta)$ obtained by setting $\alpha=0$ do not involve the reflection operator. The conclusions of our analysis will prevail also for this subfamily and, as in the case $\alpha\ne 0$, all these systems will be shown to be superintegrable. 

When $k=1$, $H_1$ has the following expression in Cartesian coordinates $(x_1=r\cos \theta, x_2=r \sin \theta)$ : 
\bel{23} H_1 =-\partial _{x_1}^2-\partial_{x_2}^2+\omega^2(x_1^2+x_2^2) +\frac{\beta^2}{4x_1^2}+\frac{\alpha^2}{4x_2^2}-\frac1{2\sqrt{x_1^2+x_2^2}}\left(\frac{\alpha x_1}{x_2^2}R_2+\frac{\beta x_2}{x_1^2}\right),\ee where $R_2$ is the reflection operator in $x_2$.
Now let 
\bel{24} \phi=k\theta, \qquad -\frac{\pi}{2}\leq \phi \leq \frac{\pi}2,\ee
we then have 
\bel{25} H_k=-\partial_r^2-\frac 1r \partial_r+\omega^2r^2+\frac{k^2}{r^2}H_\phi,\ee
where 
\bel{26} H_\phi=-\partial_\phi^2+\frac{\alpha k^2}{2}\left(\frac{\frac\alpha 2-\cos \phi\ R}{\sin^2\phi}\right)+\frac{\beta k^2}{2}\left(\frac{\frac\beta 2-\sin \phi }{\cos^2\phi}\right).\ee

The angular operator $H_\phi$ corresponds (up to a factor) to the exactly solvable 1-d Hamiltonian that was studied in \cite{18}. It is manifestly an extension of the original Scarf I Hamiltonian \cite{20}. It was shown to be supersymmetric within a framework where supercharges are taken to be differential-difference operators with reflections\cite{P1, 18}. In the present case, it is readily checked that 
\bel{27} H_\phi=Q_{\alpha, \beta}^2\ee
 with 
 \bel{28}Q_{\alpha, \beta}=\left(\partial_\phi-\frac{\beta}{2\cos \phi}\right)R-\frac{\alpha}{2\sin \phi}.\ee
 Obviously, the Hamiltonians $H_k$ have been constructed by combining the radial part of the 2-d harmonic oscillator Schr\"odinger operator with this supersymmetric 1-d Hamiltonian. $H_\phi$ is thus set as a constant of motion and $H_k$ is therefore integrable (at a minimum). The exact solvability of the eigenvalue equations associated to $H_\phi$ ensures that the 2-d Hamiltonian also shares that property through separation of variables as we shall see next. 
\section{Exact solvability}
We are looking for the solutions $\Psi(r,\phi)$ to the Schr\"odinger equation 
\bel{31} H_k\Psi(r, \phi)=E\Psi(r,\phi)\ee
with $H_k$ as in \eref{25} and \eref{26}. (The solutions to the original problem are obtained by making the substitution \eref{24}.) Let us separate the variables and write 
\bel{32} \Psi(r,\phi)=\Omega(r) \Phi(\phi).\ee
The separated equations read 
\be \label{33} \left(-\partial_r^2-\frac{1}{r}\partial_r-\frac{ka^2}{r^2}\right)\Omega(r)=E\Omega(r).\ee
and \be \label{34} H_\phi\Phi(\phi)=a^2\Phi(\phi)\ee
where $H_\phi$ and $Q_{\alpha, \beta}$ are as in \eref{26} and \eref{27} respectively. 

The solutions to the radial equation can be obtained by taking 
\bel{36} \Omega(r)=r^\gamma e^{-\omega^2 r^2/2}F(r). \ee
The resulting equation for $F$ is 
\bel{37} \left[ -\partial_r^2+\left(2\omega r+\frac{(2\gamma+1)}{r}\right)\partial_r +\frac{k^2a^2-\gamma^2}{r^2}+2\omega(\gamma+1)\right]F=EF.\ee
Choose 
\bel{38} \gamma=k|a|\ee 
and set \bel{39} E=\mathcal{E} +2\omega(k|a|+1)\ee
to find 
\bel{310} \left[ -\partial_r^2+\left(2\omega r+\frac{(2k|a|+1)}{r}\right)\partial_r\right] =\mathcal{E}F.\ee  
Performing the change of variable 
\bel{311} y=\omega r^2,\ee
\eref{310} becomes
\bel{312}y\partial_y^2F+(1+k|a|-y)\partial_yF+\frac{\mathcal{E}}{4\omega}F=0\ee
which is recognized \cite{21} as the equation that has the Laguerre polynomials $L_m^{k|a|}$ as regular solutions if 
\bel{313} \mathcal{E}=4\omega m. \ee
The radial part of the wave functions which are continuous at the origin for $\omega >0$ and vanishing at infinity are thus given by 
\bel{314} \Omega(r)=M_mY^{k|a|}_m(y),\ee
where 
\be Y^{k|a|}_m(y)= (y)^{k|a|/2} e^{-y/2}L_m^{k|a|}(y)\ee
and $M_m$ is a normalization constant that we shall specify later. From \eref{39} and \eref{313} the eigenvalues are 
\bel{315} E_m=2\omega[2m+k|a| +1].\ee

Let us know turn to the angular equation \eref{34}. This equation has been solved in \cite{18} by obtaining the eigenfunctions $\Phi$ of $Q_{\alpha, \beta}:$
\bel{316} Q_{\alpha, \beta}\Phi(\phi)=a\Phi(\phi).\ee
Clearly, $H_\phi\Phi=a^2\Phi$. One first observes that 
\bel{317} \Phi_0=N_0 |\sin \phi|^{\alpha/2}\cos^{\beta/2}\phi(1+\sin \phi)^{1/2}\ee
satisfies 
\bel{318} Q_{\alpha, \beta}\Phi_0=-\frac12(\alpha+\beta+1)\Phi_0.\ee
One then carries out the "gauge transformation" of $Q_{\acb}$ with $\Phi_0$: 
\bel{319} \widetilde{Q}_{\acb} =\Phi_0^{-1}Q_{\acb}\Phi_0; \qquad \Phi=\Phi_0G\ee
and after performing the change of variable 
\bel{320} x=\sin \phi \ee
one finds that the functions $G$ obey the equation 
\bel{321}  \widetilde{Q}_{\acb}G=aG\ee
with 
\bel{322} \widetilde{Q}_{\acb}=(1-x)\partial_x R-\frac{\alpha}{2x}(1-R)-\frac{\apb+1}2R.\ee
This equation \eref{322} is recognized to be the differential-difference equation \cite{14, 18}
that has the little -1 Jacobi polynomials $\pab$ as regular solutions. The eigenvalues are 
\bel{323} a_n=\left\{ \ba{cc} -n -\frac{\apb +1}2& \qquad \mbox{ n even } \\
n +\frac{\apb +1}2& \qquad \mbox{ n odd } \ea \right. . \ee
These polynomials have the following expressions in terms of hypergeometric terminating series 
\bel{325} \fl P_{n}^{(\alpha, \beta)}(x)=\chi_n\left[{}_2F_1\left(\ba{cc} -\frac n2 &\frac{n+\alpha +\beta+2}2\\ &\frac{\alpha+1}2 \ea; x^2\right)+\frac{nx}{\alpha+1}{}_2F_1\left(\ba{cc} 1-\frac n2 &\frac{n+\alpha +\beta+2}2\\ &\frac{\alpha+3}2 \ea; x^2\right)\right]\ee
for $n$ even; and 
\bel{326} \fl  P_{n}^{(\alpha, \beta)}(x)=\chi_n\left[{}_2F_1\left(\ba{cc} \frac{1-n}2 &\frac{n+\alpha +\beta+1}2\\ &\frac{\alpha+1}2 \ea ; x^2\right)-\frac{(\alpha+\beta+1)x}{\alpha+1}{}_2F_1\left(\ba{cc} \frac{1-n}2 &\frac{n+\alpha +\beta+3}2\\ &\frac{\alpha+3}2 \ea; x^2\right)\right]\ee
for $n $ odd. 

Recall that the series 
\bel{326a} {}_2F_1\left(\ba{cc} a, & b\\ & c\ea; x\right)=\sum_{n=0}^\infty \frac{(a)_n(b)_n}{n! (c)_n}x^n\ee
terminates whenever $a$ or $b$ is a negative integer. Here $(a)_n=a(a+1)\cdots(a+n-1)$ etc. are shifted factorials. 

The coefficients $\chi_n$ are chosen so as to make the polynomials $\pab$ monic, i.e. $\pab=x^n +\mathcal{O}(n-1)$. They are given by 
\bel{327}\chi_n=\left\{ \ba{cc} 
(-1)^{\frac n2} \frac{\left(\frac {\alpha+1} 2\right)_{\frac n 2 }}{\left(\frac n 2 +\frac \alpha 2 +\frac\beta 2+1\right)_{\frac n 2}}& \qquad \mbox{  n even } \\
(-1)^{\frac{n+1}{2}} \frac{\left(\frac {\alpha+1} 2\right)_{\frac {n+1} 2 }}{\left(\frac {n+1} 2 +\frac \alpha 2 +\frac\beta 2+1\right)_{\frac {n+1} 2}}& \qquad \mbox{  n odd } \ea .\right. \ee
For $\alpha >-1$ and $\beta>-1,$ the polynomial $\pab$ are orthogonal with respect to the weight function 
\bel{328}   \omega(x)=|x|^{\alpha}(1-x^2)^{(\beta+1)/2}(1+x). \ee
The solutions to the angular equation \eref{34} are hence given by 
\bel{329}  \Phi_n(\phi)=\frac{N_n}{N_0}X_n(\sin\phi),\ee
with\be X_n(\sin \phi ) =\Phi_{0}(x) P_{n}^{(\alpha, \beta)}(\sin \phi)\ee
and where $N_n$ are also normalization constants that we shall specify below. Finally, the eigensolutions $\Psi$ of \eref{31} are 
\bea\label{330}\fl \Psi_{n,m}&=&\frac{M_{m} N_n}{N_0}Y^{k|a_n|}_m(y)X_n(x)\\
\fl&=& \frac{M_{m} N_n}{N_0}(\sqrt{\omega}r)^{k|a_n|}e^{-\omega^2r^2/2}|\sin \phi|^{\alpha/2}\cos^{\beta/2}\phi(1+\sin \phi)^{1/2}L_m^{k|a_n|}(\omega r^2)P_{n}^{(\alpha, \beta)}(\sin \phi)\nonumber\eea
and the corresponding spectra are 
\bel{331} E_{m,n}=2\omega\left[2m +k\left(n+\frac{\apb+1}2\right)+1\right].\ee
Very much like the in the case of the TTW systems, the energy eigenvalues are linear in the quantum numbers $n$ and $m.$ For rational $k$, there is a degeneracy of states that we shall explain in the next section. 

The inner product on the space of wave functions is naturally defined by 
\bel{332} 
\langle f(r, \theta), g(r, \theta)\rangle =\int_0^\infty r dr\ \int_{-\frac \pi2}^{\frac{\pi}2}d\phi\ \overline{f(r,\phi)} g(r,\phi).\ee
It can then be checked that the eigenfunctions $\Psi_{m,n}$ of $H_k$ (with $k$ fixed) are orthonormal
\bea \label{333}\fl \langle \Psi_{m,n}, \Psi_{m',n'}\rangle&=&\frac{M_{m}M_{m'}N_{n'}N_n}{2\omega N_0^2 }\int_0^\infty dy Y^{k|a_n|}_m(y)Y^{k|a_n'|}_{m'}(y) \left(X_n(x), X_{n'}(x)\right)\\
\fl &=&\delta_{m,m'}\delta_{n,n'}\nonumber\eea
where 
\bel{334} \left(X_n(x), X_{n'}(x)\right)=\int_{-1}^{1}dx \sqrt{1-x^2}X_n(x) X_{n'}(x)\ee
is the inner product on the space of wave functions of $H_\phi$. Owing to the orthogonality of the polynomials $P_n^{\acb}(x)$ with respect to the weight function \eref{328}, the functions $X_n(x)$ are orthogonal. The norms have been evaluated in \cite{18}.  Recall $x=\sin \phi$, and we have
\bea \fl  \left(X_n(x), X_{n'}(x)\right)&=&\int_{-\frac \pi2}^{\frac \pi2}d(\sin \phi)|\sin \phi|^\alpha (\cos\phi)^{\beta-1}(1+\sin\phi)P_{n}^{\acb}(\sin \phi) P_{n'}^{\acb}(\sin \phi)\nn
\fl &=&\int_{-1}^1 dx \ |x|^\alpha (1-x^2)^{\frac{\beta-1}2}(1+x)P_{n}^{\acb}(x) P_{n'}^{\acb}(x)\nn
\fl &=&\delta_{n,n'}\frac{N_0^2}{N_n^2} \eea
The normalization constants $N_n$ are given by \cite{18}:
\be \label{Nn} \fl N_n=\left\{ \ba{lc}
 \frac{N_0\left(\frac \alpha 2+\frac \beta 2+1\right)_{n} }{\sqrt{ \left({\frac n2}\right)!\left(\frac \alpha 2 +\frac\beta 2+1\right)_{\frac n2}\left(\frac\alpha 2+\frac 12\right)_{\frac n2}\left(\frac \beta 2+\frac 12\right)_{\frac n2}}}, &\quad  \mbox{ for $n$ even,} \\
\frac{N_0\left(\frac \alpha 2+\frac \beta 2+1\right)_{n} }{\sqrt{ \left({\frac n2-\frac12}\right)!\left(\frac \alpha 2 +\frac\beta 2+1\right)_{\frac n2-\frac12}\left(\frac\alpha 2+\frac 12\right)_{\frac n2+\frac12}\left(\frac \beta 2+\frac 12\right)_{\frac n2+\frac12}}},&\quad  \mbox{ for $n$ odd,} \\
\ea \right.\ee
with
\bel{336}N_0=\left[\frac{\Gamma\left(\frac\alpha 2+\frac\beta 2 +1\right)}{\Gamma\left(\frac \alpha 2+1)\right)\Gamma\left(\frac \beta 2+1)\right)}\right]^{1/2},\ee
and where $\Gamma(x)$ is the standard gamma function. As a result 
\bea \fl \langle \Psi_{n,m}, \Psi_{n',m'}\rangle &=&\frac{N_nN_{n'}M_{m}M_{m'}}{2\omega N_0^2}\int_0^\infty dy Y^{k|a_n|}_m(y)Y^{k|a_{n'}|}_{m'}(y) \left(X_n(x), X_{n'}(x)\right)\\
\fl \label{337} &=& \delta_{n,n'} \frac{M_{m}M_{m'}}{2\omega}\int_0^\infty dy Y^{k|a_n|}_m(y)Y^{k|a_{n}|}_{m'}(y)\\
\fl \label{338} &=&\delta_{n,n'} \frac{M_{m}M_{m'}}{2\omega} \int_0^\infty dy\ (y)^{k|a_n|} e^{-y}L_m^{k|a_n|}(y) L_{m'}^{k|a_{n}|}(y)\\
\fl &=&\delta_{n,n'}\delta_{m,m'} \eea
by the orthogonality of the Laguerre polynomials and assuming that 
\bel{339} M_{m,n}=\sqrt{\frac{2\omega m!}{\Gamma(m+k|a_n|+1)}},\ee
where we have appended the index $n$ to show the dependence on both $m$ and $n$. 

\section{Superintegrability}
We now want to prove that for rational values of $k$, the Hamiltonians $H_k$ are maximally superintegrable  in addition to being exactly solvable. To that end, we shall explicitly identify the additional constants of motion that are at the root of the state degeneracies. We shall also determine the defining relations of the symmetry algebra  formed by these conserved quantities. The supercharge $Q_{\acb} $ will be found to play an important role. To proceed, we shall use the recurrence relation approach that has been developed in \cite{4}. This method uses the structural properties possessed by the separated eigenfunctions, to obtain the constants of motion from appropriate combinations of the corresponding ladder operators. For the Laguerre polynomials, these relations are known \cite{21}, as for the little -1 Jacobi polynomials, they have been determined in \cite{17}. 

We look for recurrence relations on the wave functions that leave the energy fixed, i.e. automorphisms on the energy eigenspaces.   To this end, assume $k=p/q$, then the transformations 
\bea \label{mpmp}  m \rightarrow m+p, \qquad n\rightarrow n-2q\\ m\rightarrow m-p, \qquad n\rightarrow n+2q \eea
do not change the energy:
\be E=\frac{2\omega}{q}\left(2mq+pn+p(\alpha+\beta +1)+q\right).\ee
Recall, that the wave functions are indexed by $m,\ n$ as
\[ \Psi_{n,m}=\frac{N_nM_{m,n}}{N_0}Y_{m}^{k|a_n|}(y)X_n(x), \qquad y=\omega r^2, \quad x=\sin \phi,\]
with 
\[ Y^{k|a_n|}_m(y)= (y)^{k|a_n|/2} e^{-y/2}L_m^{k|a_n|}(y)\]
\[ X_n(x) =\Phi_{0}(x) P_{n}^{(\alpha, \beta)}(x).\] 
\subsection{Ladder operators for the radial part}
Based on the ladder operators for the Laguerre polynomials \cite{21}, we have the following ladder operators for the functions $Y_m^{k|a_n|}(y)$
\bea \fl K_{k|a_n|,E}Y_m^{k|a_n|}(y)&=\left[ (1+k|a_n|)\partial_y-\frac{E}{4\omega}-\frac{1}{2y}k|a_n|(1+k|a_n|)\right]Y_m^{k|a_n|}(y)\\
\fl &=-Y_{m-1}^{k|a_n|+2}(y)\\
\fl K_{-k|a_n|,E}Y_m^{k|a_n|}(y)&=\left[ (1-k|a_n|)\partial_y-\frac{E}{4\omega}+\frac{1}{2y}k|a_n|(1-k|a_n|)\right]Y_m^{k|a_n|}(y)\\
\fl &=-(m+1)(m+k|a_n|) Y_{m+1}^{k|a_n|-2}(y),\eea
where $ E=E_{m,n}=2\omega[2m+k|a_n| +1].$
%Note that the change in $k|a_n|$ gives a change in $n$
%\[ k|a_n|-2=ka_n-2=\frac{p}q(n-2\frac{q}p+\frac{\alpha+\beta+1}{2})=ka_{n-2\frac{q}{p}}.\]
To obtain the desired shift in the quantum numbers \eref{mpmp} we take a composition of $K$'s  $p$ times, with the corresponding value of $k|a_n|$ shifted in each successive application
\bea K^{p}_{k|a_n|,E}\equiv K_{k|a_n|+2(p-1),E}\cdots K_{k|a_n|+2,E} K_{k|a_n|, E}\\
     K^{p}_{-k|a_n|,E}\equiv K_{-(k|a_n|-2(p-1)),E}\cdots K_{-(k|a_n|-2),E} K_{-k|a_n|,E}.\eea
The action of these operators on the wave functions is given by 
\bea K^{p}_{k|a_n|,E}Y_{m}^{k|a_n|}X_n=(-1)^pY_{m-p}^{k|a_n|+2p}X_n\\
     K^{p}_{-k|a_n|,E}Y_{m}^{k|a_n|}X_n=(-1)^p(m+1)_p(k|a_n|+m-p+1)_p Y_{m+p}^{k|a_n|-2p}X_n.\eea
It is important to note that although the quantity $E$ is a function of $m$ and $n$, it is unchanged by the operation $m\rightarrow m\pm 1, $ and $k|a_n| \rightarrow k|a_n| \mp 2.$ Hence, the change does not need to be accounted for in the successive applications of the operator. 

\subsection{Ladder operators for the angular part}
Next, we look for operators which will change the value $n$ by $\pm 2q.$ To obtain these, we use the ladder operators of the little $-1$ Jacobi polynomials, as given in \cite{17}. The operators are:
\bea J_+=(x+(1-x)R)\left(2\widetilde{Q}_{\acb}+1\right)+\alpha+\beta\nn
J_-=(x-(1-x)R)\left(2\widetilde{Q}_{\acb}-1\right)+\alpha-\beta\eea
It is straightforward to verify that these operators satisfy 
\be\label{JpmQ}  \lbrace J_+,\widetilde{Q}_{\acb}\rbrace =-J_+, \qquad \lbrace J_-, \widetilde{Q}_{\acb}\rbrace =J_-\ee
and so the operators $J_+$ and $J_-$ fix the space of $P_n$ but change the degree of the polynomials by plus or minus 1, depending on the parity of the degree, $n$. The action of these operators on the polynomials $P_n$ is
\bea J_+P_n&=\Bigg\{\ba{lc} -\frac{2n(\alpha+\beta+n)}{(\alpha+\beta+2n)}P_{n-1}  &\mbox{for n  even,}\\
2(\alpha+\beta+2n+2) P_{n+1} &\mbox{ for n odd,}\ea\\
 J_-P_n&=\Bigg\{\ba{lc} -2(\alpha+\beta+2n+2) P_{n+1}  &\mbox{for n  even,}\\
\frac{2(\alpha+n)(\beta+n)}{(\alpha+\beta+2n)}P_{n-1} & \mbox{ for n odd.} \ea.\eea
The operator which will change the degree of  $X_n$ by 2q is given by 
\be\label{jm2q} J^{-,2q}\equiv \Phi_0 (J_-J_+)^q\Phi_0^{-1}X_n=\Bigg\{\ba{lc} c_n X_{n-2q}  &\mbox{for n  even,} \\
                                                                   d_n X_{n+2q}  & \mbox{for n  odd,} .\ea \ee 
where 
\bea  c_n=\frac{(-2^4)^q\left(-\frac n 2\right)_q \left(-\frac {\alpha+\beta+n} 2\right)_q\left(\frac {1-\alpha-n} 2\right)_q\left(\frac {1-\beta-n} 2\right)_q}{\left(-\frac {\alpha+\beta} 2-n\right)_{2q}},\\
  d_n= (-2^4)^q\left(\frac{\alpha+\beta}2 +n+1\right)_{2q}. \eea 
Writing the $J$'s in the the opposite order, we obtain 
\be \label{jp2q} J^{+,2q}\equiv \Phi_0 (J_+J_-)^q\Phi_0^{-1}X_n=\Bigg\{\ba{lc} \tilde{d}_n X_{n+2q}  & \mbox{for n  even,} \\
                                                                   \tilde{c}_n X_{n-2q}  & \mbox{for n  odd,}\ea \ee 
where 
\bea  \tilde{c_n}=\frac{(-2^4)^q\left(-\frac {n-1} 2\right)_q \left(-\frac {\alpha+\beta+n-1} 2\right)_q\left(-\frac {\alpha+n} 2\right)_q\left(-\frac {\beta+n} 2\right)_q}{\left(-\frac {\alpha+\beta} 2-n\right)_{2q}},\\
  \tilde{d}_n= d_n. \eea 
Thus, we have operators with the desired action on the basis states, however, their action depends on the parity of the quantum number $n.$

\subsection{Parameter independent ladder operators}
Next, we would like to combine the ladder operators for the radial and angular parts to obtain operators which give automorphism of the energy eigenspace and which are independent of the indices $m$ and $n$.  On the one hande, note that while the action of the operators $J^{\pm,2q}$ depend on the parity $n,$ the operators do not. On the other hand, the operators $K^{p}_{\pm k|a_n|,E}$ depend on the values of $n$ and $m$. To remove this dependence, we shall replace $|a_n|$ by the operator $Q_{\alpha,\beta}$. This has the added advantage of making the action of the $K^\pm$ depend on the parity of $n$. Thus, let us define new operators which are now differential operators in $x$ and $y$ with a reflection in $x$:
\bea \label{kQ} K_{kQ,E}\equiv \left[ (1+kQ_{\alpha,\beta})\partial_y-\frac{E}{4\omega}-\frac{1}{2y}kQ_{\acb}(1+kQ_{\alpha,\beta})\right]\nn
     K_{kQ,E}Y_{m}^{k|a_n|}X_n=\Bigg\{ \ba{cc}  K_{-k|a_n|,E}Y_{m}^{k|a_n|}X_n, &\mbox{ for n  even,}\\
K_{k|a_n|,E}Y_{m}^{k|a_n|}X_n, &\mbox{ for n  odd.} \ea
\eea
The repeated application of this operator, denoted  $K_{kQ,E}^p,$ is defined as
\be  K_{kQ,E}^p=K_{kQ+2(p-1),E}\cdots K_{kQ+2,E} K_{kQ,E}\ee
and will act on the basis as 
\be K_{kQ,E}^pY_{m}^{k|a_n|}X_n=\Bigg\{ \ba{cc}  K_{-k|a_n|,E}^pY_{m}^{k|a_n|}X_n &\mbox{ for n  even,}\\
                                                      K_{k|a_n|,E}^pY_{m}^{k|a_n|}X_n &\mbox{ for n  odd.} \ea.\ee Similarly, 
\be  K_{-kQ,E}^pY_{m}^{k|a_n|}X_n=\Bigg\{ \ba{cc}   K_{k|a_n|,E}^pY_{m}^{k|a_n|}X_n &\mbox{ for n  even,}\\
                                                     K_{-k|a_n|,E}^pY_{m}^{k|a_n|}X_n &\mbox{ for n  odd.} \ea.\ee
Note that the operator $K_{kQ,E}$ still has $m$ and $n$ dependence via the energy $E $ and so the next step is to remove the energy from the operator. To do this, we push the constant $E$ to the right and replace it with the Hamiltonian as
\be \label{kQH} K_{kQ,H}^p=K_{kQ,E}^p|_{E=H}, \qquad K_{-kQ,H}^p=K_{-Q,E}^p|_{E=H},.\ee
This operator,  $K_{kQ,H}^p$ is parameter independent and acts on  the basis by 
\bea\fl K_{kQ,H}^pY_{m}^{k|a_n|}X_n=K_{kQ,E}^pY_{m}^{k|a_n|}X_n\\
\fl\quad   = \Bigg\{ \ba{lr}   K_{-k|a_n|,E}^pY_{m}^{k|a_n|}X_n=(-1)^p(m+1)_p(k|a_n|+m-p+1)_p Y_{m+p}^{k|a_n|-2p}X_n &\mbox{ for n  even,} \\
                                                      K_{k|a_n|,E}^pY_{m}^{k|a_n|}X_n=(-1)^pY_{m-p}^{k|a_n|+2p}X_n &\mbox{ for n  odd.} \ea\nonumber \eea

Hence, we have parameter independent operators $\Xi_1$ and $\Xi_2$ given by 
\be \label{xidef} \Xi_1\equiv \Phi_0(J_+J_-)^q\Phi_0^{-1}K_{kQ,H}^p, \qquad \Xi_2\equiv \Phi_0(J_-J_+)^q\Phi_0^{-1}K_{-kQ,H}^p,\ee
which act on the basis as 
 \bea  \Xi_1 Y_{m}^{k|a_n|}X_n =\Bigg\{\ba{lc}  \ell^-_{m,n}Y_{m+p}^{k|a_n|-2p}X_{n-2q} &\mbox{ for n  even,}\\
\ell_{m,n}^{+}Y_{m-p}^{k|a_n|+2p}X_{n +2q} &\mbox{ for n  odd,} \ea\\
\Xi_2 Y_{m}^{k|a_n|}X_n =\Bigg\{\ba{lc}  \tilde{\ell}_{m,n}^{+}Y_{m-p}^{k|a_n|+2p}X_{n+2q} &\mbox{ for n  even,}\\
\tilde{\ell}_{m,n}^{-}Y_{m+p}^{k|a_n|-2p}X_{n-2q} &\mbox{ for n odd,} \ea\eea 
with
\bea\fl\label{lm}  \ell_{m,n}^{-}\! \!=\!\frac{(-2^4)^q(-1)^p(m+1)_p(k|a_n|\!+m-p+1)_p\!\!\left(-\frac n 2\right)_q\! \!\left(-\frac {\alpha+\beta+n} 2\right)_q\!\!\left(\frac {1-\alpha-n} 2\right)_q\!\left(\frac {1-\beta-n} 2\right)_q}{\left(-\frac {\alpha+\beta} 2-n\right)_{2q}},\\
\fl \label{tlm} \tilde{\ell}_{m,n}^{-}\! \!=\!\frac{(-2^4)^q(-1)^p(m+1)_p(k|a_n|\!+m-p+1)_p\!\!\left(\frac {1-n} 2\right)_q \!\!\left(\frac {1-\alpha-\beta-n} 2\right)_q\!\!\left(-\frac {\alpha+n} 2\right)_q\!\!\left(-\frac {\beta+n} 2\right)_q}{\left(-\frac {\alpha+\beta} 2-n\right)_{2q}}, \\
\fl\label{lp}  \ell_{m,n}^{+} =\tilde{\ell}_{m,n}^{+} =(-1)^p (-2^ 4)^q\left(\frac{\alpha+\beta}2 +n+1\right)_{2q} .
\eea
From the action on the wave functions, we can use a standard Wronskian argument as in \cite{4} to show that the operators $\Xi_1$ and $\Xi_2$ commute with the Hamiltonians. However, it is also possible to prove this explicitly as we have done in the appendix. 

Also from the action on the basis, we can verify that the operators $\Xi_1$ and $\Xi_2$ are mutual adjoints. Since the action of $\Xi_1$ and $\Xi_2$ on the basis involve only one term, it is enough to show that, for $n$ even 
\bel{adje} \langle \Psi_{m+p, n-2q}, \Xi_1 \Psi_{m,n}\rangle=\langle \Xi_2 \Psi_{m+p, n-2q}, \Psi_{m,n} \rangle\ee
and, for $n$ odd, 
\bel{adjo} \langle \Psi_{m-p, n+2q}, \Xi_1 \Psi_{m,n}\rangle =\langle \Xi_2\Psi_{m-p,n+2q}, \Psi_{m,n}\rangle. \ee
For $n$ even, the requirement \eref{adje} is equivalent to 
\be \frac{N_{n}^2M_{m,n}^2  }{N_{n-2q}^2M_{m+p, n-2q}^2}=\frac{\tilde{\ell^+}_{m+p,n-2q}}{\ell^-_{m,n}}.\ee
This can be directly verified since
\bea \fl \frac{N_{n}^2M_{m,n}^2  }{N_{n-2q}^2M_{m+p, n-2q}^2}&=\frac{\left(-\left(\frac{\alpha+\beta}2\right)-n\right)_{2q}^2}{\left(-\frac n2\right)q\left(-\frac{\alpha+\beta+n}2\right)_q\left(\frac{1-\alpha-n}2\right)_q\left(\frac{1-\beta-n}2\right)_q(m+1)_p(m-p+k|a_n|+1)_p}\nn
&=\frac{\tilde{\ell^+}_{m+p,n-2q}}{\ell^-_{m,n}}.
\eea
For $n$ odd, the requirement \eref{adjo} is equivalent to 
\be \frac{N_{n}^2M_{m,n}^2  }{N_{n+2q}^2M_{m-p, n+2q}^2}=\frac{\tilde{\ell^-}_{m-p,n+2q}}{\ell^+_{m,n}}.\ee
Which again is directly verified as 
\bea \fl \frac{N_{n}^2M_{m,n}^2  }{N_{n+2q}^2M_{m-p, n+2q}^2}&= \frac{m!\left(\frac{n+1}2\right)_q\left(\frac{\alpha+\beta+1+n}2\right)_q\left(\frac{\alpha+n}2+1\right)_q \left(\frac{\beta+n}2+1\right)_q\left(m+k|a_n|+1\right)_p}{(m-p)!\left(n+1+\frac{\alpha+\beta}2\right)_{2q}^2}\nn
&=\frac{\tilde{\ell^-}_{m-p,n+2q}}{\ell^+_{m,n}}. \eea
Thus, $\Xi_1^\dagger=\Xi_2$ and we have a superintegrable system with $H, Q_{\alpha,\beta}$ and self-adjoint operators $\Xi_1+\Xi_2$ or $i(\Xi_1-\Xi_2).$   
\subsection{Algebra Relations}
From the expansion coefficients \eref{lm}-\eref{lp}, we can use the action of the operators on the basis $Y_{m}^{k|a_n|}X_n$ to determine the algebra satisfied by the operators $\Xi_1, \Xi_2 $ and $Q_{\acb}.$
The commutator of $\Xi_1$ with $Q_{\alpha,\beta}$ is given by
\bea \fl [\Xi_1, Q_{\alpha,\beta}]\Psi_{m,n}=\Bigg\{\ba{lc}-|a_n|\Xi_1 Y_{m}^{k|a_n|}X_n-Q_{\alpha,\beta} \ell_{m,n}^{-}Y_{m+p}^{k|a_n|-2p}X_{n-2q} &\mbox{ for n even,} \\
                                       \quad |a_n|\Xi_1 Y_{m}^{k|a_n|}X_n-Q_{\alpha,\beta} \ell_{m,n}^{+}Y_{m-p}^{k|a_n|+2p}X_{n+2q}&\mbox{ for n odd,} \ea\nn
=\Bigg\{\ba{lc}(-|a_n|+|a_{n-2q}|) \ell_{m,n}^{-}Y_{m+p}^{k|a_n|-2p}X_{n-2q} &\mbox{ for n even,}\\
                                       (|a_n|-|a_{n+2q}|) \ell_{m,n}^{+}Y_{m-p}^{k|a_n|+2p}X_{n+2q}&\mbox{ for n odd,}\ea\nn
=-2q\Xi_1\Psi_{m,n}.\eea
Similarly,
\bea \fl [\Xi_2, Q_{\alpha,\beta}]\Psi_{m,n}=\Bigg\{\ba{lc}-|a_n|\Xi_2 Y_{m}^{k|a_n|}X_n-Q_{\alpha,\beta} \tilde{\ell}_{m,n}^{+}Y_{m-p}^{k|a_n|+2p}X_{n+2q} &\mbox{ for n even,}\\
                                       |a_n|\Xi Y_{m}^{k|a_n|}X_n-Q_{\alpha,\beta} \tilde{\ell}Y_{m+p}^{k|a_n|-2p}X_{n-2q}&\mbox{ for n odd,} \ea\nn
=\Bigg\{\ba{lc}(-|a_n|+|a_{n+2q}|) \tilde{\ell}_{m,n}^{+}Y_{m-p}^{k|a_n|+2p}X_{n+2q} &\mbox{ for n even,}\\
                                       (|a_n|-|a_{n-2q}|) \tilde{\ell}_{m,n}^{-}Y_{m+p}^{k|a_n|-2p}X_{n-2q}&\mbox{ for n odd,} \ea\nn
=2q\Xi_2\Psi_{m,n}.\eea
From the action on the basis, we can conclude that 
\be [\Xi_1, Q_{\alpha,\beta}]=-2q\Xi_1, \qquad  [\Xi_2, Q_{\alpha,\beta}]=2q\Xi_2.\ee
Finally, writing the expansion coefficients in terms of $\epsilon =m+\frac{kn}2, $ so that $E=2\omega\left(2\epsilon+\frac{k(\alpha+\beta+1)}{2} +1\right)$, they become
\bea\fl \label{lme} \ell_{m,n}^{-} = (\epsilon-\frac{kn}2+1)_p\left(-\frac{k(n+\alpha+\beta+1)}{2}-\epsilon\right)_p\frac{(-2^4)^q\left(-\frac n 2\right)_q \left(-\frac {\alpha+\beta+n} 2\right)_q\left(\frac {1-\alpha-n} 2\right)_q\left(\frac {1-\beta-n} 2\right)_q}{\left(-\frac {\alpha+\beta} 2-n\right)_{2q}}\nn
\fl  \tilde{\ell}_{m,n}^{-} = (\epsilon-\frac{kn}2+1)_p\left(-\frac{k(n+\alpha+\beta+1)}{2}-\epsilon\right)_p\frac{(-2^4)^q\left(\frac {1-n} 2\right)_q \left(\frac {1-\alpha-\beta-n} 2\right)_q\left(-\frac {\alpha+n}2\right)_q\left(-\frac {\beta+n} 2\right)_q}{\left(-\frac {\alpha+\beta} 2-n\right)_{2q}}\nn
\fl \label{lpe} \ell_{m,n}^{+}=\tilde{\ell}_{m,n}^{+} =(-1)^p (-2^4)^q\left(\frac{\alpha+\beta}2 +n+1\right)_{2q}.\eea
Using these forms of the coefficients \eref{lme}-\eref{lpe}, the action of $[\Xi_1, \Xi_2]$ on the basis is:
\be\label{x1x2} [\Xi_1, \Xi_2]\Psi_{m,n}= \Bigg\{\ba{lc}(\tilde{\ell}^+_{m,n}\ell^-_{m-p,n+2q}-\tilde{\ell}^+_{m+p, n-2q}\ell^-_{m,n}) \Psi_{m,n} &\mbox{ for n even,} \\
(\tilde{\ell}^-_{m,n}\ell^+_{m+p,n-2q}-\tilde{\ell}^-_{m-p, n+2q}\ell^+_{m,n})\Psi_{m,n}&\mbox{ for n odd,} \ea \ee
where
\bea\fl \tilde{\ell}^+_{m,n}\ell^-_{m-p,n+2q}&=&(-1)^p2^{8q}\left(1-\epsilon+\frac{kn}2\right)_p\left(1+\epsilon+\frac{k(n+\alpha+\beta+1)}2\right)_p
\nn
\fl &&\times
\left(1+\frac{n}2\right)_q\left(1+\frac{\alpha+\beta+n}{2}\right)_q\left(\frac{\alpha+n-1}2\right)_q\left(\frac{\alpha+n-1}2\right)_q,\\
\fl \tilde{\ell}^+_{m+p,n-2q}\ell^-_{m,n}&=&(-1)^p2^{8q}(\epsilon-\frac{kn}2+1)_p\left(-\frac{k(n+\alpha+\beta+1)}{2}-\epsilon\right)_p\nn
\fl &&\times \left(-\frac n 2\right)_q \left(-\frac {\alpha+\beta+n} 2\right)_q\left(\frac {1-\alpha-n} 2\right)_q\left(\frac {1-\beta-n} 2\right)_q,\\
\fl {\ell}^+_{m+p,n-2q}\tilde{\ell}^-_{m,n}&=&(-1)^p2^{8q}(\epsilon-\frac{kn}2+1)_p\left(-\frac{k(n+\alpha+\beta+1)}{2}-\epsilon\right)_p\nn
\fl &&\times \left(\frac {1-n} 2\right)_q \left(\frac {1-\alpha-\beta-n} 2\right)_q\left(-\frac {\alpha+n}2\right)_q\left(-\frac {\beta+n} 2\right)_q,\\
\fl {\ell}^+_{m,n}\tilde{\ell}^-_{m-p,n+2q}&=&(-1)^p2^{8q}\left(1-\epsilon+\frac{kn}2\right)_p\left(1+\epsilon+\frac{k(n+\alpha+\beta+1)}2\right)_p
\nn
\fl &&\times\left(\frac {1+n} 2\right)_q \left(\frac {1+\alpha+\beta+n} 2\right)_q\left(1+\frac {\alpha+n}2\right)_q\left(1+\frac {\beta+n} 2\right)_q.
\eea
These coefficients are polynomial in $n$ and $\epsilon$ and so can be written as polynomials in the operators $Q_{\alpha,\beta}$ and $E$ as
\[ \epsilon=\frac{E}{4\omega}-\frac{k(\alpha+\beta+1)}{4}-\frac12,\qquad n =|a_n|-\frac{\alpha +\beta+1}2\]
which is compatible with the substitution $E=H$ and $|a_n|=-Q_{\alpha,\beta}$ for $n$ even and $|a_n|=Q_{\alpha,\beta}$ for $n$ odd. 
The commutators of the operators $\Xi_1$ and $\Xi_2$ can thus be expressed as  
\be \fl\label{x1x2e}  \left[\Xi_1, \Xi_2 \right]=\left\{ \!\!\!\!\ba{cc}\left(\tilde{\ell}^+_{m,n}\ell^-_{m-p,n+2q}-\tilde{\ell}^+_{m+p, n-2q}\ell^-_{m,n}\right)|_{ \epsilon=\frac{H-\omega k(\alpha+\beta+1)-2\omega}{4\omega}, n =-\frac{2Q_{\alpha,\beta}+\alpha +\beta+1}2} & \mbox{for n even,} \\ \left(\tilde{\ell}^-_{m,n}\ell^+_{m+p,n-2q}-\tilde{\ell}^-_{m-p, n+2q}\ell^+_{m,n}\right)|_{\epsilon=\frac{H-\omega k(\alpha+\beta+1)-2\omega}{4\omega}, n =\frac{2Q_{\alpha,\beta}-\alpha -\beta-1}2} &\mbox{for n odd.}\ea \right. \ee

To guarantee that these two forms of \eref{x1x2e} are equal, it is enough to verify that under the transformation  $ n\rightarrow -(n+\alpha+\beta+1)$ the coefficients transform as
\bea \tilde{\ell}^+_{m,n}\ell^-_{m-p,n+2q}\rightarrow \tilde{\ell}^-_{m,n}\ell^+_{m+p,n-2q},\nn
\tilde{\ell}^+_{m+p, n-2q}\ell^-_{m,n}\rightarrow \tilde{\ell}^-_{m-p, n+2q}\ell^+_{m,n}.\eea
This requirement ensures that replacing $|a_n|$ with $\pm Q_{\alpha,\beta}$ will give the appropriate expansion \eref{x1x2} for both even and odd basis and so the two forms in \eref{x1x2e}. Thus, we have shown that the commutator and analogously the anti-commutator is a polynomial, of degree at least $2p+4q$ in the operator $Q_{\alpha,\beta}$ and $2p$ in the Hamiltonian.  

\subsection{Case $k=1$}
Let us present explicitly the results  for the case $k=1$. The ladder operators are given by 
\bea\fl \Xi_1&=\left((x+\sqrt{1-x^2}R)\left(2Q_{\alpha,\beta}+2\right)+\alpha+\beta\right)\left((x-\sqrt{1-x^2}R)\left(2Q_{\alpha,\beta}-1\right)+\alpha-\beta\right)\nn
\fl &\times \left[ (1+kQ_{\alpha,\beta})\partial_y-\frac{H_1}{4\omega}-\frac{1}{2y}kQ_{\alpha,\beta}(1+Q_{\alpha,\beta})\right],\\
\fl\Xi_2&=\left((x-\sqrt{1-x^2}R)\left(2Q_{\alpha,\beta}-1\right)+\alpha-\beta\right)\left((x+\sqrt{1-x^2}R)\left(2Q_{\alpha,\beta}+1\right)+\alpha+\beta\right)\nn
\fl &\times \left[ (1-kQ_{\alpha,\beta})\partial_y-\frac{H_1}{4\omega}+\frac{1}{2y}kQ_{\alpha,\beta}(1-Q_{\alpha,\beta})\right].\eea

Finally, we can compute explicitly the structure relations for the symmetry algebra for $k=1$:
\be [\Xi_1, Q_{\alpha,\beta}]=-2\Xi_1, \qquad  [\Xi_2, Q_{\alpha,\beta}]=2\Xi_2, \ee
\bea\fl [\Xi_1,\Xi_2]=48Q_{\alpha,\beta}^4-8\frac{H_1^2+2\omega^2(\alpha^2+\beta^2-9)}{\omega^2}Q_{\alpha,\beta}^3-2\frac{3\alpha\beta}{2}Q_{\alpha,\beta}^2+\frac{\alpha\beta}{\omega^2}(H^2_1-4\omega^2)\\
\fl \qquad +\left(2\frac{\alpha^2+\beta^2-3}{\omega^2}H_1^2+(\beta-\alpha-3)(\beta+\alpha-3)(\beta-\alpha+3)(\beta+\alpha+3)-3\right)Q_{\alpha,\beta}\nonumber,\eea
and 
\bea \fl \{\Xi_1,\Xi_2\}=-8Q_{\alpha,\beta}^6+2\frac{H_1^2+2\omega^2(\alpha^2+\beta^2)-58\omega^2}{\omega^2}Q_{\alpha,\beta}^4+8\alpha\beta Q_{\alpha,\beta}^3\\
 -\frac{(2\alpha^2+2\beta^2-22)H_1+\omega^2((\alpha^2-\beta^2)^2-50(\alpha^2+\beta^2)+193}{2\omega^2}Q_{\alpha,\beta}^2\nn
-\frac{8\alpha\beta(H_1^2-12\omega^2)}{3\omega^2}Q_{\alpha,\beta}+\frac{(\alpha^2-\beta^2)^2-10(\alpha^2+\beta^2)+9}{8\omega^2}(H_1^2-4\omega).
\eea
Note that these algebra relations are degree $7$ and $8$ in the momenta, respectively. 

\section{An infinite family of superintegrable systems with $1/r$}
As was shown for the TTW system \cite{17}, coupling constant metamorphosis \cite{23, 24} can be used to map this new infinite family of superintegrable systems $H_k$ \eref{21} to a different family of superintegrable systems $\widetilde{H}_k$ with a Coulomb term, 
\be \label{th} \widetilde{H}_k=-\partial_\rho^2-\frac1\rho\partial_\rho+\frac{\gamma}{\rho}+\frac{k^2}{4\rho^2}H_{\frac{\psi}{2}}.\ee

To see this mapping, take the Schr\"odinger equation 
\bel{hke} \left(-\partial_r^2-\frac1r\partial_r+\omega^2r^2+\frac{k^2}{r^2}H_{\phi}\right)\Psi-E\Psi=0\ee
and solve for $\omega^2/4$ as
\be \left(-\frac{1}{4r^2}\partial_r^2-\frac{1}{4r^3}\partial_r-\frac{E}{4r^2}+\frac{k^2}{4r^4}H_{\phi}\right)\Psi(r,\phi)=-\frac{\omega^2}4\Psi(r,\phi).\ee
Then, making the change of variables 
\bel{r} r^2=\rho, \qquad \phi=\frac{\psi}2\ee
and interchanging the coupling constants with the energies
\bel{E} E=-4\gamma, \qquad -\frac{\omega^2}4=\widetilde{E},\ee
we obtain the Hamiltonian \eref{th} with Schr\"odinger equation
\be \label{thke} \widetilde{H}_k\widetilde{\Psi}=\widetilde{E}\widetilde{\Psi}.\ee
The wave functions in \eref{hke} and \eref{thke} differ only in the change of variables \eref{r} and parameters \eref{E}. 
The separation constant $k|a_n|$ becomes $k/2|a_n|$ and so the quantization of the spectra $ E=\omega(2m+k|a_n|+1)$ becomes 
\be-4\gamma=2\sqrt{-\tilde{E}}(2m+\frac{k}2|a_n|+1) \ee
or 
\be \tilde{E}=-\frac{8\gamma^2}{(4m+k|a_n|+2)^2}.\ee
The wave functions $\widetilde{\Psi}$ are given by 
\be \widetilde{\Psi}=\frac{N_nM_{m,n}}{N_0}Y_{m}^{\frac{k}{2}|a_n|}(\tilde{y})X_n(\tilde{x}), \qquad \tilde{y}=2\sqrt{-\widetilde{E}}\rho, \quad \tilde{x}=\sin\left(\frac{\psi}2\right) .\ee

Furthermore, as was shown in \cite{17, 24} the integrals of motion are preserved under this transformation. Alternatively, we can use the recurrence relation method described above to obtain the integrals, since the wave functions are essentially the same. Note that the integrals of $H_k$ will be mapped to integrals of $\widetilde{H}_{k/2}.$

\section{Conclusion}
In this paper, we have constructed an infinite family of exactly solvable Hamiltonians indexed by a parameter $k$ whose wave functions can be written in terms of Laguerre and little -1 Jacobi polynomials. For rational $k$, we used the recurrence relation approach of Kalnins, Kress and Miller \cite{4} to construct the integrals of motion out of the ladder operators of the two families of orthogonal polynomials. This method not only gives explicit expressions for the integrals of motion but also gives the symmetry algebra which they satisfy. 

We have thus provided one class of Hamiltonians with reflections corresponding to superintegrable systems in the plane. It would be of interest to look systematically for analogous models, that is to search for all separable Hamiltonians that are superintegrable when the occurrence of reflection operators is allowed. This would lead to the identification of orthogonal polynomials that fall outside of the Bochner Theorem in \cite{16}. 

 \ack
 The authors would like to thank W. Miller Jr.  and P. Winternitz for discussions.  The work of (LV) is supported in part through funds provided by the National Science and Engineering Research Council (NSERC) of Canada. (SP) acknowledges a postdoctoral fellowship provided by the Laboratory of Mathematical Physics of the CRM, Universit\'e de Montr\'eal. 

\appendix
\section{Direct proof of the commutation relations}\label{AppendixA}
\setcounter{section}{1}
Let us prove explicitly that the operator $\Xi_1$ commutes with $H_k$. The proof for $\Xi_2$ is analogous. 
\begin{theorem}
The operator $\Xi_1$ \eref{xidef} commutes with the Hamiltonian $H_k$ \eref{21} for $k=p/q$.
\end{theorem}

To this end, we prove by induction the following lemmas:
\begin{lemma} 
The following identity holds for all $\ell:$
\[ [ J^{+,2\ell}, H_k]=-J^{+,2\ell}\frac{4\omega k^2(\ell^2-\ell Q_{\acb})}{y}.\]
\end{lemma}
{\bf Proof}: This is a proof by induction. Recall the definition \eref{jp2q}
\be J^{+,2\ell}=\Phi_0(J_+J_-)^\ell\Phi_0^{-1}.\ee
As a direct result of the anti-commutation relations of the operators $J_\pm$  \eref{JpmQ},
we obtain
\bea \left[ J^{+,2}, Q_{\alpha,\beta} \right]=2 J^{+,2}\\
 \label{al1} \left[  J^{+,2}, Q_{\alpha,\beta}^2 \right]=-4 J^{+,2}(1-Q_{\acb}).\eea
Next, we use the form of the Hamiltonian \eref{21} in terms of the variable $y=\omega r^2:$
\be\label{hy} H_k=\omega\left(-4y\partial_y^2-4\partial_y+y+\frac{k^2Q^2_{\acb}}y\right).\ee
And so, since $J_\pm$ have no dependence on $y$, their commutator with $H_k$ is given by 
\be [ J^{+,2\ell}, H_k]=\frac{\omega k^2}{y} [J^{+,2\ell}, Q^2_{\acb}].\ee
The final step is then to prove by induction that 
\be \label{al}  \left[J^{+,2\ell}, Q^2_{\acb}\right]=-4J^{+,2\ell}(\ell^2-\ell Q_{\acb}).\ee
The case $\ell=1$ is shown in \eref{al1} and so we assume \eref{al} for $\ell-1$ and compute 
\bea \fl \left[J^{+,2\ell}, Q^2_{\acb}\right]&=&J^{+,2\ell-2}\left[J^{+,2}, Q^2_{\acb}\right]+\left[J^{+,2\ell-2}, Q^2_{\acb}\right]J^{+,2}\\
&=&-4J^{+,2\ell}(1-Q_{\acb})-4J^{+,2\ell-2}\left((\ell-1)^2-(\ell-1) Q_{\acb}\right)J^{+,2}\\
&=&-4J^{+,2\ell}(\ell^2-\ell Q_{\acb}).\eea
Thus, by induction we have proved the lemma. 
\qed

Next, we will prove an identity for the commutator of the operator $ K_{kQ,H}^p$ with $H_k. $ Along the way, we will give an explicit factorized form for the operator $ K_{kQ,H}^p.$ 

\begin{lemma}
The following identity holds for all $\ell:$
\be \label{indk} [ K_{kQ,H}^\ell, H_k]=\frac{4\omega}{y}\left(\ell^2+\ell kQ\right)K_{kQ,H}^\ell.\ee
\end{lemma}
{\bf Proof:} First, we note that for $\ell=1$, the operator $K_{kQ,H}^1$ is
\be\label{kkqh1} K_{kQ,H}^1= \left[ (1+kQ_{\alpha,\beta})\partial_y-\frac{H_k}{4\omega}-\frac{1}{2y}kQ_{\alpha,\beta}(1+kQ_{\alpha,\beta})\right].\ee
It is then  straightforward to verify, using the explicit forms of the operators \eref{hy} and \eref{kkqh1}, that \eref{indk} holds for $\ell=1:$
\be \left[K_{kQ,H}^1, H_k\right]=\frac{4\omega}{y}\left(1+kQ\right)K_{kQ,H}^1.\ee
We shall prove the lemma again using induction, so assume \eref{indk} for $\ell-1.$
For the purposes of induction, we take the factorized form of $K_{kQ,H}^\ell$ as in \eref{kQ}
\be K_{kQ,H}^\ell=K_{kQ+2\ell-2,H}K_{kQ,H}^{\ell-1}\ee
where $K_{kQ+2\ell-2,H}$ is defined  as the operator with action on the basis as 
\be K_{kQ+2\ell-2,H}K_{kQ,H}^{\ell-1}\Psi=K_{kQ+2\ell-2,E}K_{kQ,E}^{\ell-1}\Psi,\ee
i.e. we move the constant $E$ to the right and then make the replacement with $H$. Explicitly, we have 
\bea\fl  K_{kQ+2\ell-2,H}K_{kQ,H}^{\ell-1}=-\frac{1}{4\omega}\left[ K_{kQ,H}^{\ell-1}, H_k\right]\\
\fl \qquad + \left[ (kQ_{\alpha,\beta}+2\ell-1)\partial_y-\frac{H_k}{4\omega}-\frac{1}{2y}(kQ_{\alpha,\beta}+2\ell-2)(kQ_{\alpha,\beta}+2\ell-1)\right]K_{kQ,H}^{\ell-1}.\nn
\eea
If we use the induction assumption, \eref{indk}, then the final term $\left[ K_{kQ,H}^{\ell-1}, H_k\right]$ is proportional to $K_{kQ,H}^{\ell-1}$ and we obtain an explicit expression for $K_{kQ+2\ell-2,H}$ as
\be\label{kqlh} \fl K_{kQ+2\ell-2,H}=(kQ_{\alpha,\beta}+2\ell-1)\partial_y-\frac{H_k}{4\omega}-\frac{k^2Q^2_{\acb}-k(6\ell-5)Q_{\acb}-2(3\ell-2)(\ell-1)}{2y}.\ee
Using this expression for $K_{kQ+2\ell-2,H}$, we  directly compute 
\be \fl \left[K_{kQ+2\ell-2,H},H_k\right]=4\omega\left[\frac{\ell^2+\ell k Q}{y}K_{kQ+2\ell-2,H}-K_{kQ+2\ell-2,H}\frac{(\ell-1)^2+(\ell-1)kQ_{\acb}}{y}\right]\ee
which in turn shows 
\bea\fl  \left[K_{kQ,H}^\ell, H_k\right] &=& K_{kQ+2\ell-2,H}\left[K_{kQ,H}^{\ell-1}, H_k\right]+\left[K_{kQ+2\ell-2,H}, H_k\right]K_{kQ,H}^{\ell-1}\nn
&=&\frac{4\omega(\ell^2+\ell k Q)}{y} K_{kQ,H}^\ell.\eea
and so we have proven by induction that the expression \eref{indk} holds. \qed

{\bf Proof of Theorem 1} The proof of theorem immediately follows from the two lemmas. If $k=p/q$ then, 
\bea \fl \left[ \Xi_1, H_k\right]&=& \left[J^{+,2q}K_{kQ,H}^p,H_k\right]\nn
\fl &=&  J^{+,2q}\left[K_{kQ,H}^p,H_k\right]+ \left[J^{+,2q},H_k\right]K_{kQ,H}^p\nn
   \fl &=& J^{+,2q}\left(\frac{4\omega(p^2+\frac{p^2}{q}Q)}{y}\right)K_{kQ,H}^p -J^{+,2q}\left(\frac{4\omega p^2(q^2-q Q_{\acb})}{q^2y}\right)K_{kQ,H}^p\nn
\fl &=&0.\eea
\qed

\end{document}